\documentclass[aip,pop,numerical,preprint,amsmath,amssymb]{revtex4-1}

\usepackage{natbib}
\usepackage[colorlinks=true,linkcolor=red,anchorcolor=blue,citecolor=green]{hyperref}
\usepackage[pdftex]{graphicx}
\DeclareGraphicsExtensions{.pdf,.jpeg,.png}
\usepackage{subfigure}
\graphicspath{{figures/}}
\usepackage{epstopdf}
\begin{document}
\bibcite{Lazarian}{{1}{2015}{{Lazarian\ \emph  {et~al.}}}{{Lazarian, Eyink, Vishniac,\ and\ Kowal}}}
\bibcite{snyder}{{2}{2011}{{{Snyder}\ \emph  {et~al.}}}{{{Snyder}, {Groebner}, {Hughes}, {Osborne}, {Beurskens}, {Leonard}, {Wilson},\ and\ {Xu}}}}
\bibcite{firstdw}{{3}{1964}{{Galeev, Moiseev,\ and\ Sagdeev}}{{}}}
\bibcite{bscott}{{4}{1997}{{Scott}}{{}}}
\bibcite{firstemdw1}{{5}{1972}{{Mikhailovsky}}{{}}}
\bibcite{firstemdw2}{{6}{1992}{{Mikhaĭlovskiĭ\ and\ Laing}}{{}}}
\bibcite{mirnov04a}{{7}{2004}{{Mirnov, Hegna,\ and\ Prager}}{{}}}
\bibcite{sauppe}{{8}{2016}{{Mirnov\ \emph  {et~al.}}}{{Mirnov, Sauppe, Hegna,\ and\ Sovinec}}}
\bibcite{SOVINEC2004}{{9}{2004}{{Sovinec\ \emph  {et~al.}}}{{Sovinec, Glasser, Gianakon, Barnes, Nebel, Kruger, Schnack, Plimpton, Tarditi,\ and\ Chu}}}
\bibcite{Ellis}{{10}{1980}{{Ellis, Mardenmarshall,\ and\ Majeski}}{{}}}
\bibcite{freidbergbook}{{11}{1987}{{Freidberg}}{{}}}
\bibcite{Bellan2004Fundamentals}{{12}{2004}{{Bellan}}{{}}}
	\title{Two-fluid MHD Regime of Resistive Drift-Wave Instability}
	\author{Shangchuan Yang}
        \affiliation{CAS Key Laboratory of Geospace Environment and Department of Engineering and Applied Physics, University of Science and Technology of China, Hefei, Anhui 230026, China}
        \author{Ping Zhu}
	\email{pzhu@ustc.edu.cn}
        \affiliation{CAS Key Laboratory of Geospace Environment and Department of Engineering and Applied Physics, University of Science and Technology of China, Hefei, Anhui 230026, China}
        \affiliation{KTX Laboratory and Department of Engineering and Applied Physics, University of Science and Technology of China, Hefei, Anhui 230026, China}
        \affiliation{Department of Engineering Physics, University of Wisconsin-Madison, Madison, Wisconsin 53706, USA}
	\author{Jinlin Xie}
	\affiliation{CAS Key Laboratory of Geospace Environment and Department of Engineering and Applied Physics, University of Science and Technology of China, Hefei, Anhui 230026, China}
        \affiliation{KTX Laboratory and Department of Engineering and Applied Physics, University of Science and Technology of China, Hefei, Anhui 230026, China}
	\author{Wandong Liu}
	\affiliation{KTX Laboratory and Department of Engineering and Applied Physics, University of Science and Technology of China, Hefei, Anhui 230026, China}
	\date{\today}
	\begin{abstract}
		Drift instabilities contribute to the formation of edge turbulence and zonal flows, and thus the anomalous transport in tokamaks. Experiments often found micro-scale turbulence strongly coupled with large-scale magnetohydrodynamic (MHD) processes, whereas a general framework has been lacking that can cover both regimes, in particular, their coupling. In this paper, the linear resistive drift wave instability (DWI) is investigated using a full 2-fluid MHD model, as well as its numerical implementation in NIMROD code. Both analytical and numerical analyses reveal a macro-scale global drift wave eigenmode coupled with MHD dynamics and illustrate a non-monotonic dispersion relation with respect to both perpendicular and parallel wavenumbers. NIMROD results also reveal an edge-localized behavior in the radial mode structure as the azimuthal mode number increases, implying the dependence of the 2-fluid effects due to the inhomogeneous density profile. The edge-localization introduces a non-trivial dependence of the effective perpendicular wavenumber to the perpendicular mode number, which may explain the quantitative difference between the global dispersion relation and its local approximation from the conventional local theory.
	\end{abstract}
	\maketitle
	\section{Introduction}\label{sec:introduction}
	Micro-scale turbulence, such as those driven by the drift wave instability (DWI), is believed to be one of the dominant transport mechanisms in tokamak plasmas. Whereas the conventional theory focuses only on small-scale dynamics, experiments often find regimes where micro-scale turbulences are strongly coupled with large-scale MHD processes. For example, it has long been argued that turbulence can enhance the rate of magnetic reconnection\cite{Lazarian}; in fusion system, micro-turbulence due to KBM are often thought to pose constraints on the pedestal height, which in turn influences the edge localized mode\cite{snyder}. Although there have been theories that are well developed for micro-turbulence and MHD separately in their own regimes, a general theoretical framework has been missing that can simultaneously cover regimes of both micro-turbulence and MHD processes, as well as their coupling.

	In early DWI studies, such as in A.A. Galeev et al.\cite{firstdw}, the two-fluid equations with electro-static approximation were often adopted. The magnetic fluctuation was later included in B. Scott's work\cite{bscott}. Using this model, B. Scott\cite{bscott} has investigated the electron drift turbulence in tokamak edge plasmas during L$-$H transition. These approaches adopt momentum equations for both electrons and ions, bringing an advantage in describing two-fluid effects. However, the two-fluid representation has not taken the temporal evolution of the magnetic field into consideration, and the equilibrium considered may be inconsistent with MHD force balance.

	In comparison, the single-fluid representation models DWI adopts an extended set of MHD equations, with two-fluid effects included in the generalized Ohm's law. It self-consistently includes all components of the magnetic perturbation, while still keeping both the macro-scale MHD and the dominant micro-scale 2-fluid physics. The fundamental difference between single-fluid and two-fluid representations is the treatment of the magnetic field. In single-fluid representation, the magnetic field is intrinsically included in the induction equation, which in turn guarantees MHD force balance. On the other hand, the two-fluid representation treats magnetic field only as a correction to the electrostatic approximation, such that both MHD equilibrium or self-consistent magnetic field evolution is absent. In situations where the magnetic field and its evolution are of interest or importance, the single-fluid representation might serve as a more suitable theory framework.

	The single-fluid representation has been adopted in a few DWI studies, such as those by A.B. Mikhailovskii\cite{firstemdw1,firstemdw2}, where the validity condition for the electrostatic approximation is derived. Following the same approach, V.V. Mirnov et al.\cite{mirnov04a} have derived a general theory for drift-tearing modes that cover arbitrary $\beta$ regime by taking into account plasma compressibility and the Hall term. Recently, following Mikhailovskii's study\cite{firstemdw1,firstemdw2}, a two-fluid MHD model has been adopted to develop a theory for the resistive drift instability in a plasma slab\cite{sauppe}. However, it is not clear if its conclusion would be readily applicable to the more realistic geometries and configurations since no global or geometric effects are included in slab model.
                
	In this work, the dispersion relation for DWI from the full two-fluid MHD model is first extended to a general form that is independent of the choice of coordinates. Such a coordinate-free form of dispersion relation avoids limitations and complications due to explicit dependence on coordinates, thus signifying the relevant physics. More importantly, the coordinate-free approach might be more amenable to cylindrical and toroidal geometries. Applying this model, we further compare the analytical result with the numerical result obtained from the initial-value extended full MHD code NIMROD\citep{SOVINEC2004}. Both analytical and numerical calculations in our work qualitatively agree on the essential features of linear DWI in the cylindrical geometry, which sets up the stage for the future study of the non-linear cross-scale coupling between micro-turbulences and MHD processes.

	The rest of this paper is organized as follows. The geometrical and physical configurations are described in Sec. \ref{sec:configuration}. In Sec. \ref{sec:theory}, we extend the dispersion relation for DWI in the full two-fluid MHD model, and evaluate new resistive DWI dispersion relation in cylindrical configuration. Sec. \ref{sec:calculation} reports numerical results from NIMROD calculations, along with verifications and benchmarks. Finally, summary and discussion are given in Sec. \ref{sec:summary}.

        \section{Geometry and MHD Equilibrium}\label{sec:configuration}
\begin{figure}[!htb]
		\centering
		\subfigure[``rectangular" grid]{
		\includegraphics[width=0.45\textwidth]{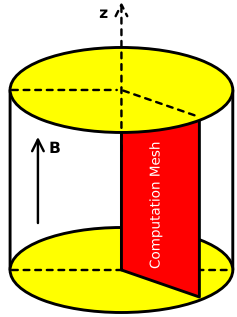}}
		\subfigure[``circular" grid]{
		\includegraphics[width=0.45\textwidth]{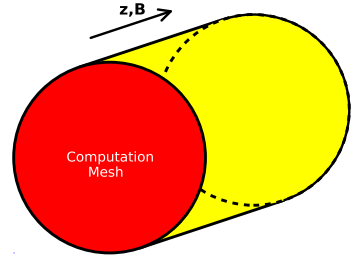}}
		\caption{\label{fig:geometry} A sketch of two different grid setups used in the NIMROD calculations for this work.}
	\end{figure}

\begin{figure}[!htb]
		\includegraphics[width=0.9\textwidth]{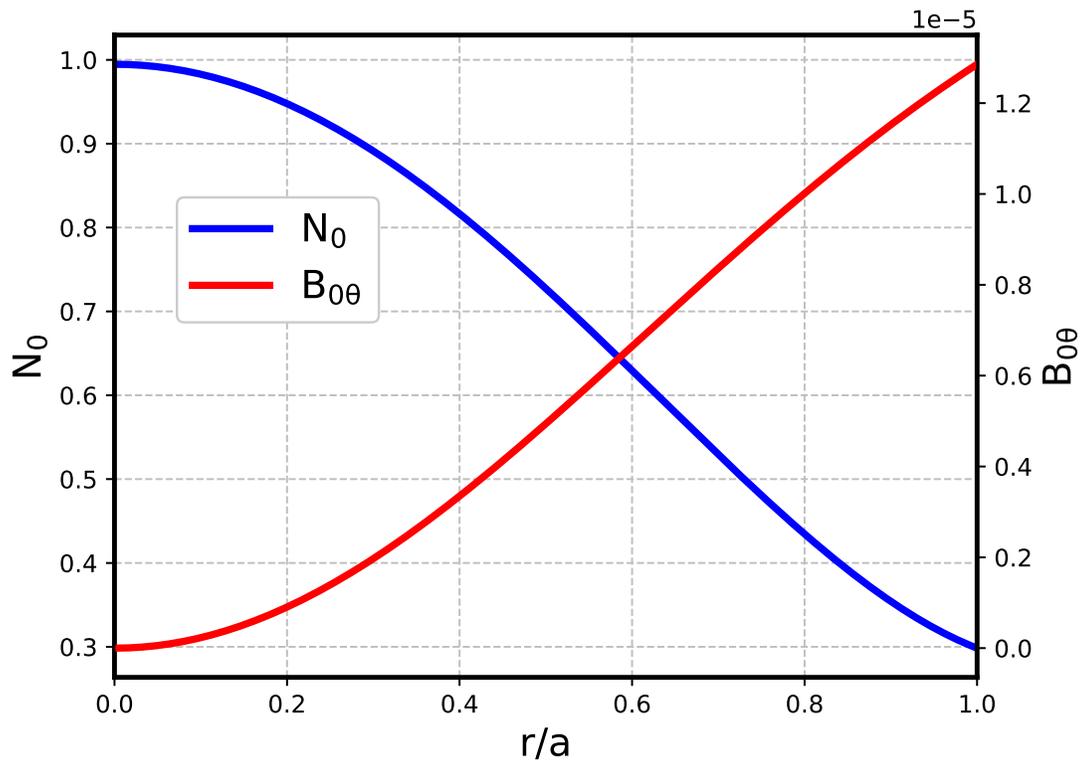}
		\caption{\label{fig:profiles}Equilibrium profiles as functions of the normalized radius $r/a$ for density $N_0$ (blue) and azimuthal magnetic field $B_{0\theta}$ (red).}
	\end{figure}

        In order to include global effects that almost always accompany macro-scale MHD processes, the cylindrical geometry is adopted for our analytical and numerical calculations. For the sake of simplicity, the ion is assumed to be cold, and the electron temperature profile uniform and constant($T_i=0$, $T_e(r)\equiv T_0$). Furthermore, the MHD equilibrium is assumed to be static. The plasma density profile $N$ is a bell-shaped function of the cylinder radius $r$, and $r$ only ($N=N(r)$). The magnetic field $\mathbf{B}$ consists of a uniform part and a non-uniform part, i.e. $\mathbf{B}\equiv\mathbf{B}_u+\mathbf{B}_e(r)$. $\mathbf{B}_u$ goes along the axial direction, and is uniform all over the cylinder. $\mathbf{B}_e$ has no radial component, either. It has a fixed pitch angle (the angle between $\mathbf{B}_e$ and axial direction), and balances the density gradient in MHD force balance equation. Additionally, $|\mathbf{B}_e|<<|\mathbf{B}_u|$ is considered such that $\beta<<1$. The geometry and MHD equilibrium profiles are illustrated in Fig. \ref{fig:geometry} and Fig. \ref{fig:profiles}, respectively.

	In comparison to more realistic configurations such as tokamaks, the equilibrium of the system has an additional direction of symmetry, i.e. the equilibrium is independent of both axial and azimuthal directions. This additional symmetry allows for an analytical formulation of DWI that is coordinate-free. The values of plasma parameters used in this work are listed in Table. \ref{tab:1}. For benchmark purpose, two kinds of grids for cylindrical geometry, namely ``circular" and ``rectangular", have been set up for NIMROD calculations (Fig. \ref{fig:geometry}). Calculations results from the ``circular" and the ``rectangular" grids have been compared to verify their correctness.

	\section{Resistive DWI under full two-fluid MHD model}\label{sec:theory}
	The two-fluid MHD model adopted in this work is based on the full set of resistive MHD equations and the inclusion of two-fluid effects in the generalized Ohm's law. Since the temperature or pressure effects are not our primary concern, a constant temperature and the equation of state $p=NT$ are assumed.

	After normalization, the rest of the full two-fluid MHD equations are:
	{
	  \color{black}
	  \begin{align}
	    \frac{\partial \hat{N}}{\partial t} = -\nabla \cdot (\hat{N} \hat{\mathbf{u}}),& \label{eq:dens} \\
	    s_1 \hat{N} (\frac{\partial \hat{\mathbf{u}}}{\partial t} + \hat{\mathbf{u}} \cdot \nabla \hat{\mathbf{u}}) = \hat{\mathbf{J}} \times \hat{\mathbf{B}} - \nabla \hat{N},& \label{eq:velo} \\
	    \frac{\partial \hat{\mathbf{B}}}{\partial t} = -\nabla \times [-\hat{\mathbf{u}} \times \hat{\mathbf{B}} + s_2 \hat{\mathbf{J}} + s_3 (\hat{\mathbf{J}} \times \hat{\mathbf{B}} - \nabla \hat{N}) + s_4 \frac{\partial \hat{\mathbf{J}}}{\partial t}],& \label{eq:magn}\\
	    \hat{\mathbf{J}} = s_1 \nabla \times \hat{\mathbf{B}},\ \ \ \ \ \ \ \ \nabla\cdot \hat{\mathbf{B}} = 0,& \label{eq:curr}
	  \end{align}
	}
	\color{black}
	For any physical quantity $F$, we denote its normalization unit and normalized form as $\bar{F}$ and $\hat{F}$, respectively. In particular, lengths are normalized in the cylinder radius $a$, times in the Alfven time $\tau_A$, velocity $\mathbf{u}$ in the Alfven speed $u_A$, and density $N$, temperature $T$, resistivity $\eta$ and magnetic field $B$ in their equilibrium peak values respectively. The detailed normalizations are given as follows
		\begin{align}
		  &u_A = \bar{B}/\sqrt{\mu_0 m_i \bar{N}},\ \ \ \ \tau_A = a/u_A,\notag\\
		  &C_s = \sqrt{\bar{T}/m_i},\ \ \ \ \bar{p} = \bar{N} \bar{T},\ \ \ \ \bar{J} = \frac{\bar{N} \bar{T}}{a \bar{B}},\notag\\
		  &\beta = \frac{\mu_0 \bar{N} \bar{T}}{\bar{B}^2},\ \ \ \ S = \frac{\mu_0 a u_A}{\bar{\eta}}\ (\text{Lundquist\ number}),\notag\\
		  &s_1 = u_A^2/C_s^2 = \beta^{-1},\notag\\
		  &s_2 = \bar{\eta} (\tau_A \bar{N} \bar{T})/(a^2 \bar{B}^2) = s_1^{-1} S^{-1},\notag\\ 
		  &s_3 = \frac{\tau_A \bar{T}}{e a^2 \bar{B}},\notag\\
		  &s_4 = \frac{m_e}{e^2} \frac{\bar{T}}{a^2 \bar{B}^2},\notag
		\end{align}
where all definitions and meanings, as well as those in Eqs.~(\ref{eq:dens})-(\ref{eq:curr}), are conventional. For an evolving field $F$, we further follow the conventional notation for linearization: $\hat{F} = F_0 + \tilde{F}$, and adopt the slow wave assumption to exclude fast MHD waves. Here a wave-like dependence $\tilde{F} \sim \exp{(2\pi inz/\bar{L}_z + i m \theta - i \omega t)}$ is assumed for the perturbation, where $m$ is the azimuthal mode number, $n$ the axial mode number, and $\bar{L}_z$ the normalized periodic length of cylinder in the $z$ direction. A set of differential equations determining the linear dispersion relation of DWI are derived using a coordinate-free approach (see Appendix) and shown as follows
{
  \color{black}
        	\begin{align}
			&{[i \omega - \frac{1}{i\omega N_0}(\mathbf{B}_0\cdot\nabla)^2 + \phi\nabla^2 + s_3N_0^{-1}(\mathbf{J}_0\cdot\nabla)](\tilde{\mathbf{B}}\cdot\nabla N_0) = s_1s_3N_0^{-1}(\mathbf{B}_0\cdot\nabla)(\mathbf{B}_0\cdot\tilde{\mathbf{B}}) }\label{eq:part1}\\
			&{[i \omega s_1 - \frac{1}{i\omega N_0}(\mathbf{B}_0\cdot\nabla)^2 + \phi\nabla^2 + s_3N_0^{-1}(\mathbf{J}_0\cdot\nabla)](\mathbf{B}_0\cdot\tilde{\mathbf{B}})}\notag\\
			&{ = [-\frac{i(\mathbf{B}_0\cdot\nabla)}{\omega N_0^2}|B_0|^2 + \frac{i}{\omega N_0 s_1}(\mathbf{B}_0\cdot\nabla) - s_1s_3N_0^{-1}(\mathbf{B}_0\cdot\nabla)(\mathbf{J}_0\cdot\nabla)](\tilde{\mathbf{B}}\cdot\nabla N_0) }\label{eq:part2}
		\end{align}
        where $\phi = s_1(s_2 \eta_0 - i\omega s_4/N_0)$.
        }        
		Effects of resistivity appear in $\phi$-dependent terms in both equations and two-fluid effects in $s_3$-dependent terms, whereas $s_1$ implies the inclusion of the coupling mechanism between MHD modes and DWI.
		Within local approximation, this dispersion relation can reduce to that in the previous work which is derived specifically for the slab configuration~\cite{sauppe}.
		The local dispersion relation for DWI is shown as follows
		\begin{align}
			&{ [\omega^2 - k_\parallel^2 + \frac{i k_\perp^2}{S}] [\omega^2 - \beta k_\parallel^2 + \frac{i\beta k_\perp^2}{S}]}\notag\\
			&{= \omega^2 \beta k_\parallel^2 k_\perp^2 d_i^2 + \omega \omega_* [\omega^2 - k_\parallel^2 + \frac{i\beta k_\perp^2}{S}] },\label{eq:disp}
		\end{align}
                where $\omega_* = - \frac{\bar{T}}{e \bar{B}} \frac{m}{r} \frac{1}{N_0}\frac{dN_0}{dr} / \tau_A$, $d_i = \frac{c}{\omega_{pi}a}$, and $k_\parallel$ and $k_\perp$ are the parallel and the perpendicular components of the wavenumber vector with respect to the equilibrium magnetic field, respectively.

\begin{figure}[!htb]
		\includegraphics[width=0.9\textwidth]{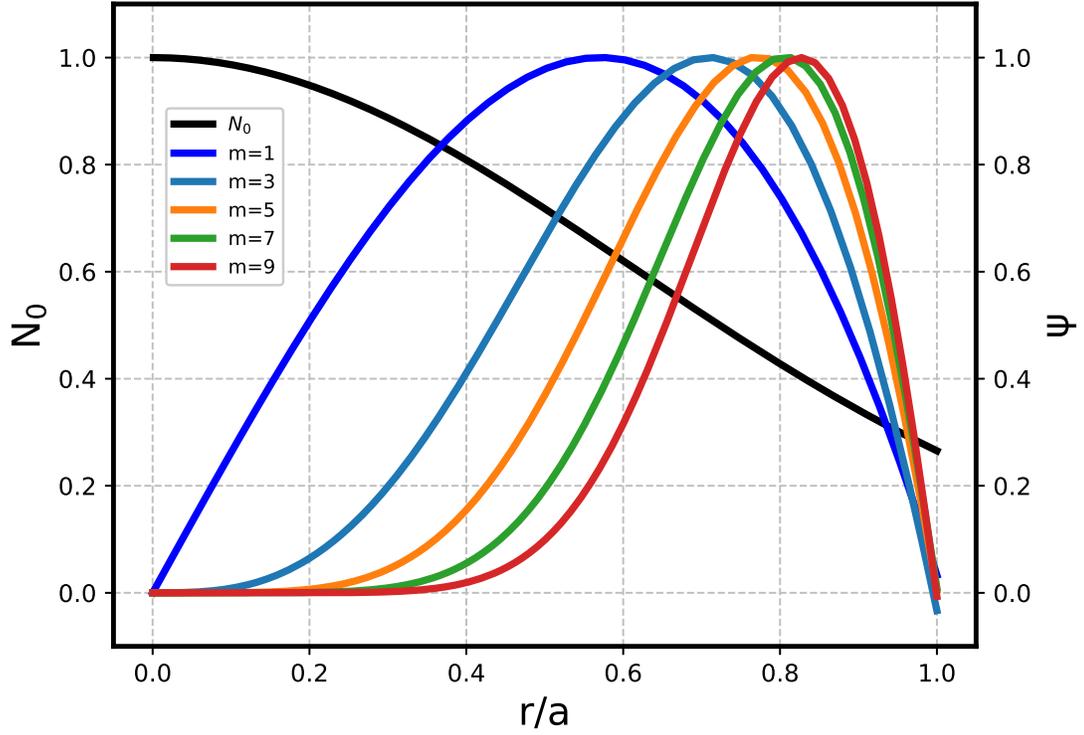}
		\caption{\label{fig:radial_str} DWI eigenmode profiles as functions of radius for different azimuthal mode number $m$ obtained from the numerical solutions of Eq.~\ref{eq:radial_struc} (colored), and the profile of the corresponding equilibrium density $N_0$ (dark).}
        \end{figure}

		To show the radial structure of the global DWI eigenmode, a simplified differential equation is derived from Eqs.~\eqref{eq:part1}-\eqref{eq:part2} in the electrostatic and $S \rightarrow \infty$ limit:
		\begin{equation}
		    \frac{d^2 \psi}{dr^2} + (\frac{1}{r} + \frac{1}{N_0}\frac{dN_0}{dr})\frac{d \psi}{dr} + (\frac{\omega_* - \omega}{\omega \rho^2} - \frac{m^2}{r^2})\psi = 0, \label{eq:radial_struc}
		\end{equation}
		where $\rho^2 = d_i^2\beta$. This result is consistent with a previous work based on the two-fluid model\citep{Ellis}. For the equilibrium considered in Sec.~\ref{sec:configuration}, the above Eq.~(\ref{eq:radial_struc}) is then solved numerically using a shooting code and the results are shown in Fig. \ref{fig:radial_str}. There the DWI eigenmode profiles are found to shift towards the cylinder edge as the azimuthal mode number $m$ increases. This behavior results in a non-trivial relation between the azimuthal mode number $m$ and the perpendicular wavenumber $k_\perp$, due to varying effective radius of DWI eigenmode for different azimuthal mode numbers.
		In fact, the concept of $k_\perp$ makes sense only if an effective radius $r_{\rm eff}$ is defined. A natural approach is to define $r_{\rm eff}$ as the radius at which the eigen-frequency of the eigenmode equals the value of local theory at $k_\perp = m / r_{\rm eff}$.
		This definition guarentees agreeement between the global theory and its local approximation, and provides a straight-forward way to calculate the value of $r_{\rm eff}$ as follows. Symbolically denote Eq.~(\ref{eq:radial_struc}) as $\frac{d}{dr} \psi(r) = \mathbf{A}(r)\psi(r)$, $r_{\rm eff}$ can be obtained by solving the local approximated dispersion relation $Det(\mathbf{A}(r;\omega,m,k_\parallel))=0$, with $m,k_\parallel$ given, and $\omega$ the corresponding eigen-frequency.

	        \section{Calculation results using NIMROD}\label{sec:calculation}

\begin{figure}[!htb]
		\includegraphics[width=0.9\textwidth]{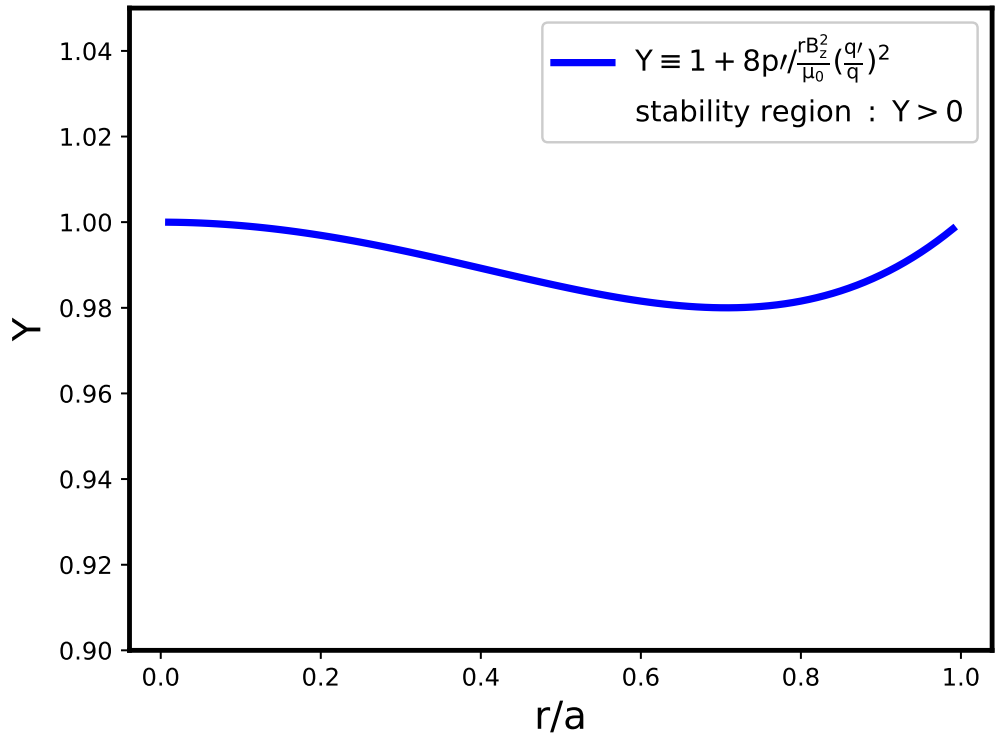}
		\caption{\label{fig:2} The Suydam's criterion parameter $Y$ as function of the normalized radius $r/a$.}
	\end{figure}

        NIMROD applies a conforming representation of 2D finite elements and 1D finite Fourier series using a semi-implicit time-advance and enables 3D nonlinear non-ideal MHD calculation in extremely stiff conditions without sacrificing the geometric flexibility needed for modeling laboratory experiments. NIMROD covers a wide range of mechanisms in tokamak plasma, including DWI, by adopting a set of full two-fluid MHD model equations as follows.

	\begin{align}
	  \mu_0 \mathbf{J} = \nabla \times \mathbf{B},\ \ \ \ \ \ \ \ \nabla\cdot \mathbf{B} = 0&,\label{eq:zzzz1}\\
	  \frac{\partial \mathbf{B}}{\partial t} = -\nabla \times \mathbf{E}&,\\
	  \frac{\partial N}{\partial t} = -\nabla \cdot (N \mathbf{u})&,\\
	  m_i N (\frac{\partial \mathbf{u}}{\partial t} + \mathbf{u} \cdot \nabla \mathbf{u}) = \mathbf{J} \times \mathbf{B} - \nabla p + \nabla\cdot\mathbf{\Pi}&,\\
	  \mathbf{E} = -\mathbf{u} \times \mathbf{B} + \eta \mathbf{J} + \frac{1}{Ne} (\mathbf{J} \times \mathbf{B} - \nabla p_e) + \frac{m_e}{Ne^2} \frac{\partial \mathbf{J}}{\partial t}&,\label{eq:zzzz6}\\
	  \frac{N_\alpha}{\gamma - 1}(\frac{\partial}{\partial t} + \mathbf{u}_\alpha \cdot \nabla)T_\alpha = -p_\alpha \nabla\cdot \mathbf{u}_\alpha - \mathbf{\Pi}_\alpha : \nabla \mathbf{u}_\alpha - \nabla \cdot \mathbf{q}_\alpha+ Q_\alpha,\ \alpha=i,e&.\label{eq:zzzz7}
	\end{align}
	The equilibrium and configuration described in Sec.~\ref{sec:configuration} are adopted in NIMROD calculations. Before carrying out further analysis, however, these configurations are first validated to ensure the domination of DWI, rather than MHD modes such as the localized interchange mode using the well-known Suydam’s criterion\cite{freidbergbook}:
        \begin{equation}
	  \frac{r B_z^2}{\mu_0}(\frac{q'}{q})^2 + 8 p' > 0\ \ \ \text{for\ stability}.\label{eq:suydam}
	\end{equation}
        Alternatively, the criterion can be rewritten as $Y\equiv 1+8 p'/\frac{r B_z^2}{\mu_0}(\frac{q'}{q})^2 > 0$ for stability.          
	For the MHD equilibrium considered here, the stability criterion is evaluated, from which the stability condition is found to be well satisfied~(Fig. \ref{fig:2}). Although Suydam's criterion (\ref{eq:suydam}) is a necessary condition for stability to the localized interchange mode, the possibility of interchange mode and other MHD modes can be further excluded by identifying the characteristic features of DWI from the calculation results below.
		
\begin{figure}[!htb]
		\centering
		\subfigure[The ``rectangular'' mesh setup]{
		\includegraphics[width=.5\textwidth]{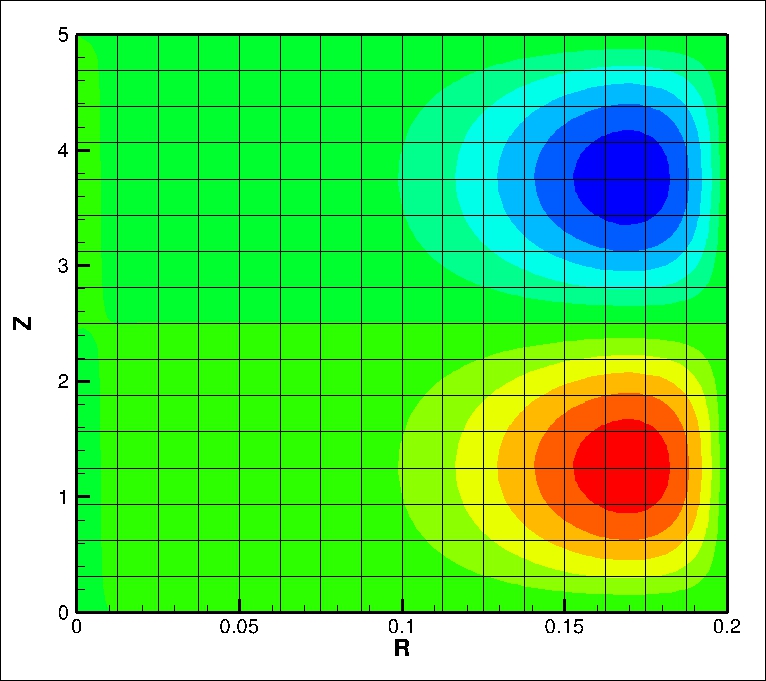}}
		\subfigure[The ``circular'' mesh setup]{
		\includegraphics[width=.5\textwidth]{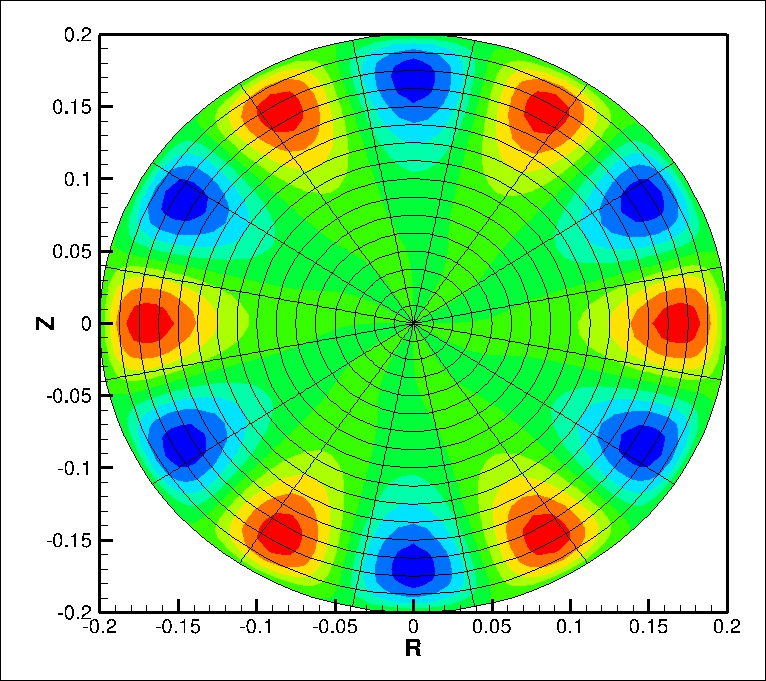}}
		\caption{\label{fig:radial_rvsc} 2D contours of $B_r$ for the $m=6,n=1$ DWI obtained from NIMROD calculations in two different mesh setups.}
	\end{figure}

	Our NIMROD calculations find a global DWI eigenmode with a bell-shaped radial profile, similar to the numerical solution of Eq.~(\ref{eq:radial_struc}). Fig. \ref{fig:radial_rvsc} illustrates one such DWI eigenmode structure for a specific set of mode numbers ($m=6,n=1$), where the calculations are successfully benchmarked between the two different computational mesh setups described in Sec.~\ref{sec:configuration}. In particular, for the same equilibrium physical parameters, the time evolutions of the DWI eigenmode yield the same growth rate from those two grids. The mode structures from the two grids are also consistent with each other (Fig. \ref{fig:radial_rvsc}).
		
	The amplitude of the eigenmode maximizes at a certain radius and vanishes at cylindrical center and edge regions. The peak value, along with its radial location, is determined globally by the MHD equilibrium profiles. This is fundamentally different from theories based on the local approximation, where only local values of MHD equilibrium fields are relevant to DWI frequency and growth rate. In situations where global effects are of interest or substantial, the difference between local and global theories may no longer be negligible.
		
\begin{figure}[!htb]
		\centering
		\subfigure[$m = 3$]{
		\includegraphics[width=.3\textwidth]{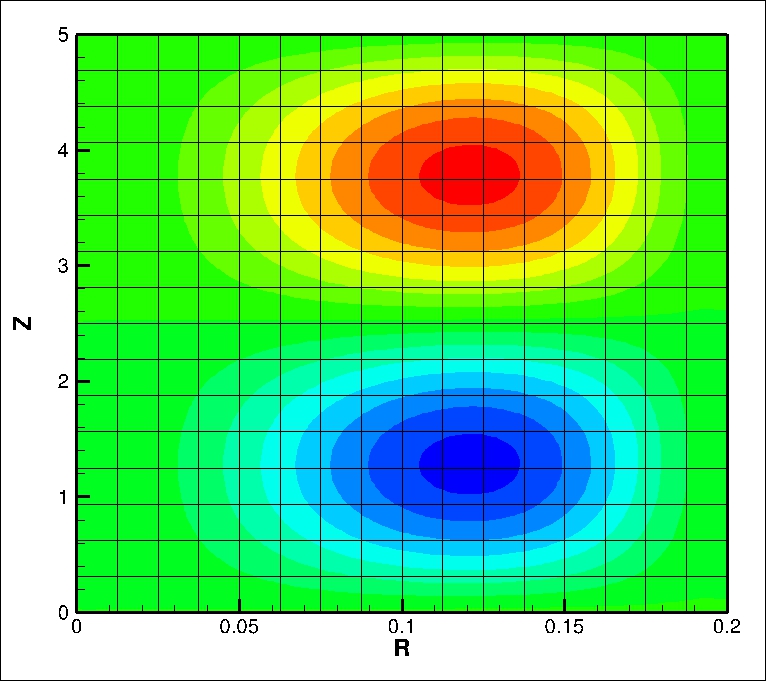}}
		\subfigure[$m = 6$]{
		\includegraphics[width=.3\textwidth]{tecplot/rect03500br.jpg}}
		\subfigure[$m = 9$]{
		\includegraphics[width=.3\textwidth]{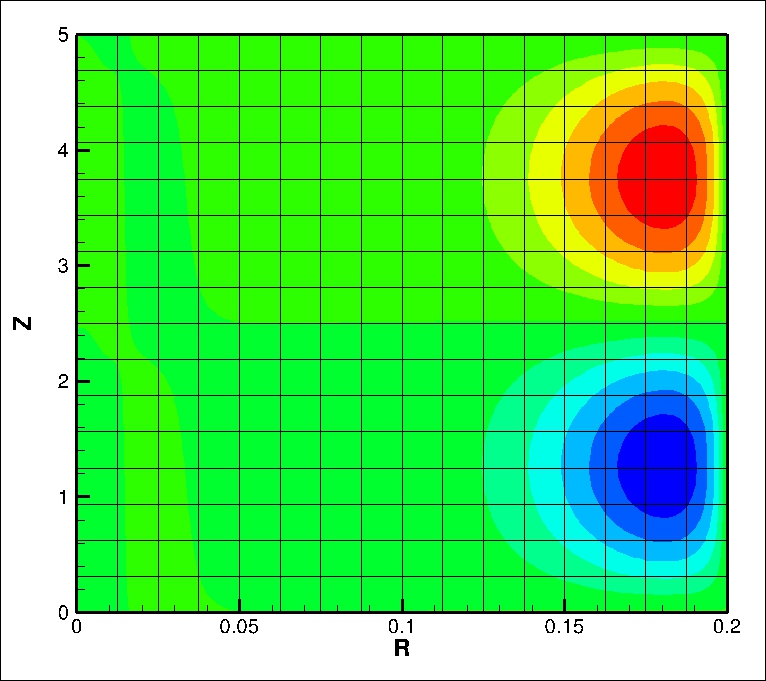}}
		\caption{\label{fig:radial_localize} 2D contours of $B_r$ for the $m=3,6,9, n=1$ drift wave instabilities.}
	\end{figure}

	Besides the DWI radial mode structure, the edge-localization of the global eigenmodes in previous analytical results are also confirmed by NIMROD results. Illustrated in Fig. \ref{fig:radial_localize} are the DWI eigenmode structures for several azimuthal mode numbers, from which the edge-localization can be clearly observed. Phenomenologically, the effective radius of DWI eigenmode $r_{\rm eff}$ becomes a function of the azimuthal wavenumber $m$, and increases as $m$ becomes larger. As a result, the effective perpendicular wavenumber $k_\perp$ depends on $m$ in a slightly more complicated manner, as in $k_\perp = m / r_{\rm eff}(m)$.
		
\begin{figure}[!htb]
		\begin{minipage}[t]{0.5\linewidth}
		\centering
		\includegraphics[width=1\textwidth]{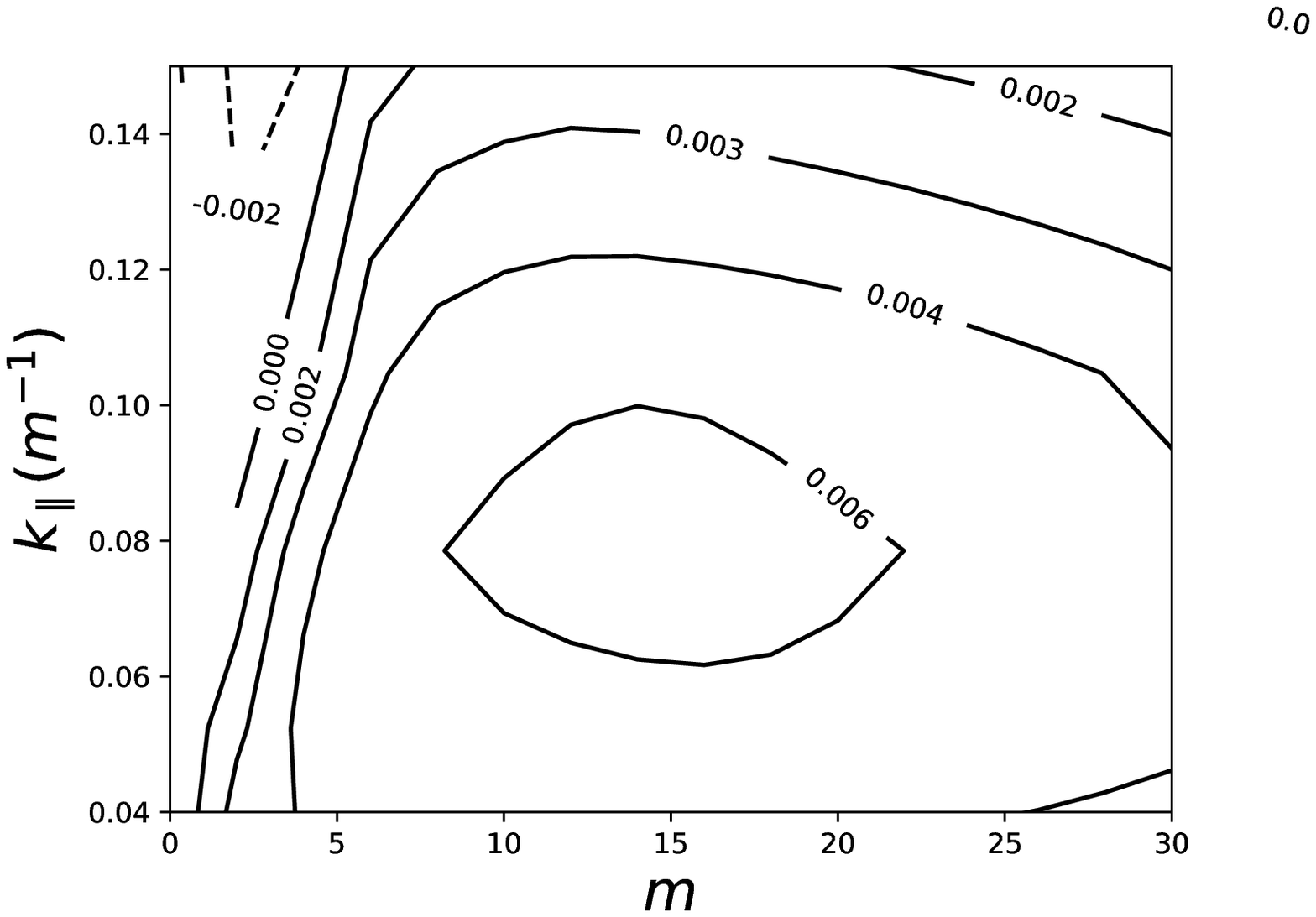}
		\end{minipage}%
		\begin{minipage}[t]{0.5\linewidth}
		\centering
		\includegraphics[width=1\textwidth]{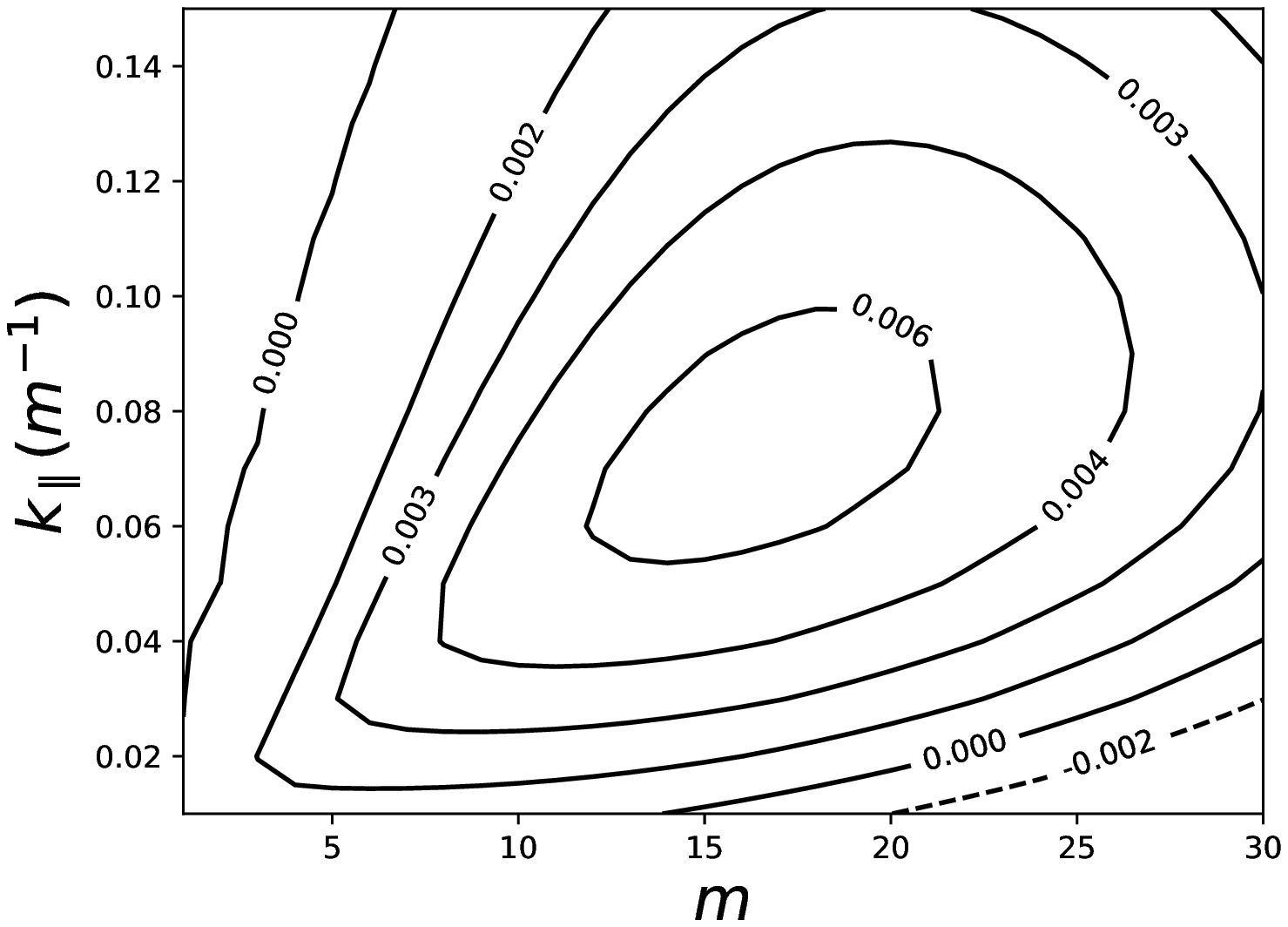}
		\end{minipage}
		\caption{Contours of linear growth rates in the perpendicular (i.e. $m$) and parallel wavenumber (i.e. $n$) plane as obtained from NIMROD calculations (Left) and analytic theory based on local approximation (Right).}\label{fig:disp}
	\end{figure}

	One of the consequences of the edge-localization appears in the dispersion relation. In particular, the linear growth rates obtained from NIMROD calculations are plotted as contours in Fig. \ref{fig:disp}, together with corresponding analytic results from the local theory in Eq.~\eqref{eq:disp}. The two results are in qualitative agreement. Both plots exhibit a non-monotonic dependence of the growth rates on the parallel and the perpendicular wavenumbers. The linear growth rate first increases with $k_\perp(k_\parallel)$, until a maximum value is reached. The quantitative difference between the two sets of contours is likely due to the global mode structure effects missing in the dispersion relation for DWI from local theory.

\begin{figure}[!htb]
		\includegraphics[width=0.9\textwidth]{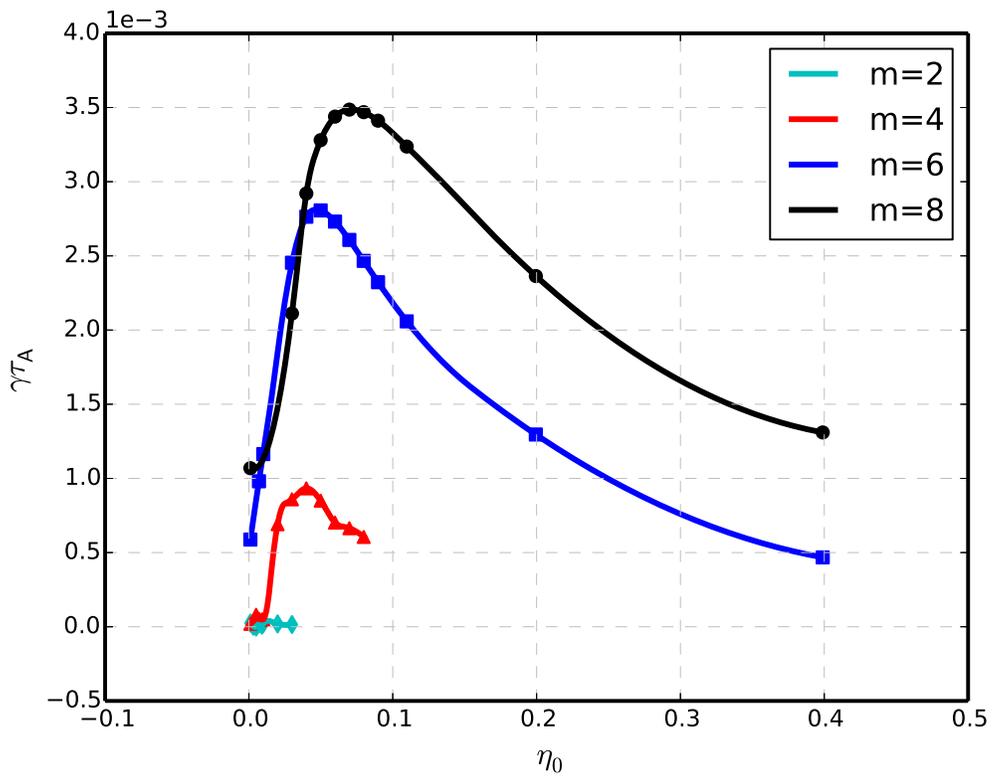}
		\caption{\label{fig:gamma_vs_eta} Linear growth rates of DWI as functions of resistivity for different azimuthal mode number $m$ as obtained from NIMROD calculations.}
	\end{figure}

        The resistive nature of the DWI eigenmode in our NIMROD calculation is confirmed from the dependence of its linear growth rate on resistivity. Fig. \ref{fig:gamma_vs_eta} shows the relation between the linear growth rate and resistivity from NIMROD results for a specific mode number. When the resistivity is small, the linear growth rate of DWI increases with resistivity, which is in agreement with the previous theories\cite{firstdw, bscott,firstemdw1,firstemdw2,Bellan2004Fundamentals}. The calculation results also find that the growth rate of DWI is reduced when the resistivity is above a certain threshold. This resistive stabilization on DWI might not have been captured by those conventional DWI theories, such as those developed by \citet{firstdw, bscott,firstemdw1,firstemdw2,Bellan2004Fundamentals}.
                
	\section{Summary, Discussion and Future Work}\label{sec:summary}
	In this work, the linear resistive drift instability is investigated using a full two-fluid MHD model, as well as its numerical implementation in NIMROD code. The analytical dispersion relation for DWI from the full two-fluid MHD model is extended to a general vector-based form that is independent of the choice of coordinates. When the eigenvalue equations for DWI are applied to cylindrical geometry, they would yield a bell-shaped radial mode structure, similar to traditional eigenmode theories on DWI. In local approximation, the eigenvalue equations reduce to the same dispersion relation in slab geometry obtained in a previous work\cite{sauppe}.
		
	Furthermore, the numerical results are obtained from the two-fluid model in NIMROD code. Our calculations have confirmed the resistive nature of the DWI eigenmode. The dispersion relation obtained from the NIMROD calculation is in qualitative agreement with our full two-fluid theory in local approximation. The comparison also shows similar radial structure profiles of DWI eigenmode from the NIMROD results and conventional eigenmode theories. In particular, the DWI eigenmode is found to localize towards the edge as the azimuthal mode number increases.

	It is worth mentioning that collision is not the only mechanism for generating DWI. In fact, the collisionless mechanism is also well-known to induce DWI. However, our work is mostly concerned about whether the full two-fluid MHD model allows the presence of DWI despite the inclusion of all components of magnetic perturbation. Such an issue should not depend on the specific mechanism, collisional or collisionless, for the excitation of DWI. The inclusion of finite resistivity prevents the cancellation of the charge separation due to fast electron response in the parallel direction, and thus serves to postpone the electron response, in a similar manner as other collisionless mechanisms.

	Although the current work is limited only to the linear regime, the analytical and numerical results in this work confirm the validity and capability of the two-fluid MHD equations for the purpose of DWI modeling. This model may provide a unified framework for future study of the nonlinear cross-scale coupling between micro-turbulences and MHD processes.

	\begin{acknowledgments}
	  This work was supported by National Magnetic Confinement Fusion Science Program of China under Grant Nos. 2014GB124002 and 2015GB101004, U.S. Department of Energy Grant Nos. DE-FG02-86ER53218 and DE-FC02-08ER54975, and the 100 Talent Program of the Chinese Academy of Sciences. This research used resources of the National Energy Research Scientific Computing Center, a DOE Office of Science User Facility supported by the Office of Science of the U.S. Department of Energy under Contract No. DE-AC02-05CH11231.
	\end{acknowledgments}

	\newpage
	\appendix
    \section{Detailed Derivation of Resistive DWI Dispersion in the Full Two-fluid MHD Model and Cylindrical Geometry}
    	In situations where good symmetry properties are present, it is often possible to represent all physics without coordinate-dependent terms.
	The rest of this appendix is the detailed derivation of the dispersion relation for global resistive DWI eigenmode in cylindrical geometry.
	The result agrees with a previous work\citep{sauppe}, where a coordinate-dependent approach is adopted.

	For the sake of brevity, the cylindrical geometry and the following are assumed: 

        1. a cold ion plasma ($T_i=0$);
        
        2. a constant temperature profile ($T_e=\bar{T}$);
        
        3. an equation of state $p=NT_e$;
        
        4. a static MHD equilibrium ($\mathbf{u}=0$).

	The normalized full two-fluid MHD equations are as follows

	    {
	    \color{black}
		\begin{align}
			&\frac{\partial \hat{N}}{\partial t} = -\nabla \cdot (\hat{N} \hat{\mathbf{u}}),\label{eq:n_n}\\
			&s_1 \hat{N} (\frac{\partial \hat{\mathbf{u}}}{\partial t} + \hat{\mathbf{u}} \cdot \nabla \hat{\mathbf{u}}) = \hat{\mathbf{J}} \times \hat{\mathbf{B}} - \nabla \hat{N},\label{eq:v_n}\\
			&\frac{\partial \hat{\mathbf{B}}}{\partial t} = -\nabla \times [-\hat{\mathbf{u}} \times \hat{\mathbf{B}} + s_2 \hat{\eta} \hat{\mathbf{J}} + s_3\hat{N}^{-1} (\hat{\mathbf{J}} \times \hat{\mathbf{B}} - \nabla \hat{N}) 
			+ s_4\hat{N}^{-1} \frac{\partial \hat{\mathbf{J}}}{\partial t}]\equiv-(\mathbf{U}+\mathbf{V}+\mathbf{W}),\label{eq:B_n}\\
			&\mathbf{U} = -\nabla\times(\hat{\mathbf{u}}\times \hat{\mathbf{B}}) = \hat{\mathbf{B}}\nabla\cdot \hat{\mathbf{u}} - \hat{\mathbf{B}}\cdot\nabla \hat{\mathbf{u}} + \hat{\mathbf{u}}\cdot\nabla \hat{\mathbf{B}},\notag\\
			&\mathbf{V} = \nabla\times[\phi/s_1 \hat{\mathbf{J}}] = \nabla\times[\phi(\nabla\times \hat{\mathbf{B}})],\notag\\
			&\mathbf{W} = s_1 s_3 \nabla\times[\frac{1}{\hat{N}}(\nabla\times\hat{\mathbf{B}})\times\hat{\mathbf{B}}],\notag\\
			&\hat{\mathbf{J}} = s_1 \nabla \times \hat{\mathbf{B}}\label{eq:J_n}.
		\end{align}
		}

    After linearization, we have:
		\begin{align}
			-i \omega \tilde{N} 	= -\nabla \cdot (N_0 \tilde{\mathbf{u}}) = -N_0\nabla\cdot\tilde{\mathbf{u}} - \nabla N_0\cdot\tilde{\mathbf{u}},&\label{eq:n}\\
			-i\omega s_1 n_0\tilde{\mathbf{u}} = (\mathbf{J}_0\times \tilde{\mathbf{B}} + \tilde{\mathbf{J}}\times \mathbf{B}_0) - \nabla\tilde{N}\label{eq:v},&\\
		    i\omega\tilde{\mathbf{B}} 	= \tilde{\mathbf{U}} + \tilde{\mathbf{V}} + \tilde{\mathbf{W}},&\label{eq:B}\\
			\tilde{\mathbf{U}} = \mathbf{B}_0\nabla\cdot\tilde{\mathbf{u}} - \mathbf{B}_0\cdot\nabla\tilde{\mathbf{u}} + \tilde{\mathbf{u}}\cdot\nabla \mathbf{B}_0,&\notag\\
			\tilde{\mathbf{V}} = - \phi\nabla^2\tilde{\mathbf{B}},& \notag\\
			\tilde{\mathbf{W}} = s_3\nabla(\frac{1}{\hat{N}})\times(\mathbf{J}_0\times \tilde{\mathbf{B}} + \tilde{\mathbf{J}}\times \mathbf{B}_0) + s_3\frac{1}{\hat{N}} \nabla\times(\mathbf{J}_0\times \tilde{\mathbf{B}} + \tilde{\mathbf{J}}\times \mathbf{B}_0),&\notag
		\end{align}

	Applying the inner products of $\nabla N_0$, $\mathbf{B}_0$ and $\nabla$ to Eq.\eqref{eq:v} respectively, we have
		\begin{align}
			\tilde{\mathbf{u}}\cdot\nabla N_0 = \frac{i}{\omega N_0}(\mathbf{B}_0\cdot\nabla)\tilde{\mathbf{B}}\cdot\nabla N_0,&\label{eq:v(n)}\\
			\tilde{\mathbf{u}}\cdot\mathbf{B}_0 = -\frac{i}{\omega N_0 s_1}\tilde{\mathbf{B}}\cdot\nabla N_0 + \frac{i}{\omega N_0}(\mathbf{B}_0\cdot\nabla)(\tilde{\mathbf{B}}\cdot\mathbf{B}_0),& \label{eq:vpara}\\
            \nabla\cdot\tilde{\mathbf{u}} = -\frac{i(\mathbf{B}_0\cdot\nabla)}{\omega N_0^2} \tilde{\mathbf{B}}\cdot\nabla N_0 - \frac{i\omega s_1}{N_0}(\tilde{\mathbf{B}}\cdot\mathbf{B}_0),&\label{eq:divv}
		\end{align}
Similarly, taking the inner products of Eq.\eqref{eq:B} with $\nabla N_0$ and $\mathbf{B}_0$ respectively yields
		\begin{align}
			i \omega(\tilde{\mathbf{B}}\cdot\nabla N_0) = -(\mathbf{B}_0\cdot\nabla)(\tilde{\mathbf{u}}\cdot\nabla N_0) - \phi\nabla^2(\tilde{\mathbf{B}}\cdot\nabla N_0) 
			+ s_3N_0^{-1} \nabla N_0\cdot\nabla\times(\mathbf{J}_0\times \tilde{\mathbf{B}} + \tilde{\mathbf{J}}\times \mathbf{B}_0),&\label{eq:part11}\\
			i \omega (\tilde{\mathbf{B}}\cdot\mathbf{B}_0) = |\mathbf{B}_0|^2\nabla\cdot\tilde{\mathbf{u}} - (\mathbf{B}_0\cdot\nabla)(\tilde{\mathbf{u}}\cdot\mathbf{B}_0) + \frac{1}{2}\tilde{\mathbf{u}}\cdot\nabla(|\mathbf{B}_0|^2)
			- \phi\nabla^2(\tilde{\mathbf{B}}\cdot\mathbf{B}_0) + \mathbf{B}_0\cdot\tilde{\mathbf{W}},&\label{eq:part22}
		\end{align}
which leads to the differential equations for the global DWI eigenmode in Eqs.~\eqref{eq:part1} and \eqref{eq:part2}.

	\newpage
	\begin{table}[!htb]
		\centering
		\begin{tabular}{l|c|r}
		\hline
		Parameter&Notation&Value\\
		\hline
		Density&$N$&$\sim2.0\times 10^{18}m^{-3}$\\
		\hline
		Electron Temperature&$T_e$&$6.25eV$\\
		\hline
		Magnetic Field&$B$&$\sim0.1T$\\
		\hline
		Cylinder Length&$Lz$&$8.0m$\\
		\hline
		Cylinder Radius&$a$&$0.1m$\\
		\hline
		Ion Cyclotron Frequency&$\omega_{ci}$&$9.58\times 10^{6}rad/s$\\
		\hline
		Ion Sound Speed&$c_s$&$2.45\times 10^{4}m/s$\\
		\hline
		Alfven Speed&$u_A$&$1.54\times 10^{6}m/s$\\
		\hline
		Ion Cyclotron Radius&$\rho_s$&$2.6mm$\\
		\hline
		Conductivity&$\eta$&$6.2\times 10^{-5}\Omega/m$\\
		\hline
		Beta&$\beta$&$5.0\times 10^{-4}$\\
		\hline
		\end{tabular}
		\caption{Typical plasma parameters used in analytical formulations and NIMROD calculations.}
		\label{tab:1}
	\end{table}
	\newpage
	\newpage
	\newpage
	\newpage
	\newpage
	\newpage
	\newpage
	\newpage
	\newpage
	
    \bibliography{bib/tex}

\end{document}